\documentclass[onecolumn]{elsart3p}
\usepackage{theorem,amssymb,epsfig}

\usepackage[english,francais]{babel}

\newtheorem{theorem}{Theorem}[section]

\newtheorem{e-proposition}[theorem]{Proposition}

\newtheorem{e-definition}[theorem]{Definition\rm}
\newtheorem{remark}{\it Remark\/}
\newtheorem{example}{\it Example\/}

\begin{document}

\centerline{ }

\begin{frontmatter}

\title{Quasicrystals, model sets, and automatic sequences}

\selectlanguage{english}
\author[authorlabel1]{Jean-Paul Allouche\thanksref{label1}}
\ead{allouche@math.jussieu.fr}
\thanks[label1]{The author was partially supported by the ANR 
project ``FAN'' (Fractals et Num\'eration).}
\author[authorlabel2]{Yves Meyer}
\ead{Yves.Meyer@cmla.ens-cachan.fr}

\address[authorlabel1]{CNRS, Institut de Math\'ematiques de Jussieu \\
\'Equipe Combinatoire et Optimisation \\
Universit\'e Pierre et Marie Curie, Case 247 \\
4 Place Jussieu, F-75252 Paris Cedex 05 (France)}
\address[authorlabel2]{CMLA, Centre de Math\'ematiques et de Leurs Applications \\
\'Ecole normale sup\'erieure de Cachan \\
94235 Cachan Cedex (France)}

\begin{abstract} 
We survey mathematical properties of quasicrystals, first from the point of view of
harmonic analysis, then from the point of view of morphic and automatic sequences.

\vskip 0.5\baselineskip

\selectlanguage{francais}
\noindent{\bf R\'esum\'e}
\vskip 0.5\baselineskip
\noindent
{\bf Quasicristaux, ensembles mod\`eles et suites automatiques. }
Nous proposons un tour d'horizon de propri\'et\'es math\'ema\-ti\-ques des quasicristaux,
d'abord du point de vue de l'analyse harmonique, ensuite du point de vue des suites
morphiques et automatiques.

\selectlanguage{english}

 \vskip 0.5\baselineskip
\noindent{\small{\it Mots-cl\'es~:} Quasicristaux~; Ensembles mod\`eles~; Suites automatiques}}
\end{abstract}
\end{frontmatter} 

\selectlanguage{english}

\section{Roots}
 
In the late sixties the second author introduced some point sets $\Lambda \subset {\mathbb R}^n$ 
which generalize lattices \cite{Mey70}, \cite{Mey72}, \cite{Mey95}, \cite{Mey12}.  
{\it Meyer sets} and {\it model sets} are defined in the next section. Roger Penrose discovered in 
1974 his famous pavings with the pentagonal symmetry.  In 1981 N.\,G.\,de Bruijn proved that the 
set $\Lambda$ of vertices of the Penrose paving is a {\it model set} (de Bruijn was unaware of 
the definition of {\it model sets} and rediscovered it). Then the diffraction pattern of $\Lambda$ 
could be computed as in \cite{Mey72}. In 1982 D.\,Shechtman discovered quasicrystals. 
D.\,Shechtman was unaware of what was achieved previously. Denis Gratias and Robert Moody 
unveiled the connections  between {\it model sets}, {\it Meyer sets} and quasicrystals.
 
\bigskip

After defining {\it model sets} and {\it Meyer sets}, we present some recent discoveries where model 
sets are playing a seminal role (Sections 3, 4, and 5).

\bigskip

Another approach to quasicrystals involves sequences generated by morphisms of the free monoid,
in particular the (binary) Fibonacci sequence. Surveying works starting from two seminal papers 
one by Kohmoto, Kadanoff, and Tang \cite{KKT}, the other by \"Ostlund, Pandit, Rand, Schellnuber,
and Siggia \cite{OPRSS}, we will describe in passing Sturmian sequences and automatic sequences. 
A last section will describe attempts to find links between the model set theory approach and 
the approach through morphic and/or automatic sequences.

\section{Model sets} 

\begin{e-definition}
A collection of points $\Lambda\subset {\mathbb R}^n$  is a Delone set if there exist two radii 
$R_2>R_1>0$ such that

\begin{itemize}

\item[(a)] every ball with radius $R_1$, whatever be its location, cannot contain more than one 
point in $\Lambda $ 

\smallskip

\item[(b)] every ball with radius $R_2$, whatever be its location, shall contain at least one point 
in $\Lambda $. 

\end{itemize}

\end{e-definition}

\begin{e-definition}
A {\it Meyer set} is a subset $\Lambda$ of ${\mathbb R}^n$ fulfilling the following
two conditions

\begin{itemize}

\smallskip

\item[(a)] $\Lambda$  is a Delone set

\smallskip

\item[(b)]  There exists a finite set $F \subset {\mathbb R}^n$  
such that  $$\Lambda \!-\!\Lambda \subset \Lambda \!+\!F.$$
\end{itemize}
\end{e-definition}

If $F=\{0\},$ then $\Lambda$ is a lattice.

\medskip

J.\,C.\,Lagarias proved in \cite{Lag99a} that (b) can be replaced by the weaker condition that $\Lambda-\Lambda$ 
is a Delone set.
\smallskip

We now define ``model sets" and unveil the connection between model sets and {\it Meyer sets}. The 
following definition can be found in \cite{Mey72}.

Let $\Gamma\subset {\mathbb R}^n\times {\mathbb R}^m={\mathbb R}^N$ be a lattice. 
If $(x, t) \in {\mathbb R}^n\times {\mathbb R}^m,$ we write $p_1(x, t)=x, \,p_2(x, t)=t.$ 
We assume that $p_1$ once restricted to $\Gamma$ is an injective mapping onto $\Gamma_1=p_1(\Gamma).$ 
We make the same assumption on $p_2.$ We furthermore assume that 
$p_1(\Gamma)$ is dense in ${\mathbb R}^n$ and $p_2(\Gamma)$ is dense in ${\mathbb R}^m$. 


\begin{e-definition}
Let $Q\subset {\mathbb R}^m$ be a compact set. Then the model set $\Lambda_Q \subset {\mathbb R}^n$ 
is defined by  
$$
\Lambda_Q=\{p_1(\gamma); \gamma \in \Gamma, \,p_2(\gamma)\in Q\}.
\leqno{(1)}$$ 
A model set is simple if $m=1$ and $Q=I$ is an interval.   
\end{e-definition}

\begin{theorem}
A model set is a {\it Meyer set}. Conversely if $\Lambda$ is a {\it Meyer set} there exists a model 
set $M$ (or a lattice $M$) and a finite set $F$ such that $\Lambda\subset M+F.$ 
\end{theorem}

\section{Almost periodic patterns}
\begin{e-definition}
A real valued function $f$ defined on ${\mathbb R}^n$ is a generalized almost periodic (g-a-p) 
function if it is a Borel function and if for every positive $\epsilon$ there exist two almost 
periodic functions $g_{\epsilon}$ and $h_{\epsilon}$ such that 
$$g_{\epsilon}\leq f \leq h_{\epsilon}\leqno{(2)}$$
and
$$ {\mathcal M}(h_{\epsilon}-g_{\epsilon})\leq \epsilon\leqno{(3)}$$
\end{e-definition}

A Borel measure $\mu$ on ${\mathbb R}^n$ is a g-a-p measure if for every compactly supported 
continuous function $g$ the convolution product $\mu*g$ is a g-a-p function. 

\begin{e-definition}
A set $\Lambda\subset {\mathbb R}^n$ is an almost periodic pattern if the associated sum of Dirac 
masses $\sigma_{\Lambda}=\sum_{\lambda\in \Lambda}\delta_{\lambda}$ is a g-a-p measure.
\end{e-definition}

\medskip

\begin{theorem}
Let us assume that the compact set $Q$ in Definition 3 is Riemann integrable. Then the corresponding 
model set $\Lambda_Q$ is an almost periodic pattern.
\end{theorem}

This is not an empty statement since almost periodic patterns have a rigid arithmetic structure as the 
following theorem shows:

\begin{theorem}
Let $\Lambda_{\theta},\,\, \theta>2,$  be the set of all finite sums $\sum_{k\geq 0}\epsilon_k\theta^k$ 
with  $\epsilon_k\in\{0, 1\}.$ Then  $\Lambda_{\theta}$ is an almost periodic pattern if and only if 
$\theta$ is a Pisot-Vijayaraghavan number. 
\end{theorem}

These two theorems are proved in \cite{Mey72}.

\section{Beyond Shannon}

The Fourier transform of $f\in L^1({\mathbb R}^n)$ will be defined by 
$$ 
\hat f(\xi)=\int_{{\mathbb R}^n} \exp(-2\pi i \xi\cdot x)f(x)dx,\,\,\, \xi\in{\mathbb R}^n.
\leqno{(4)}$$

Let $K\subset {\mathbb R}^n$ be a compact set and $E_K\subset L^2({\mathbb R}^n)$ be the translation 
invariant subspace of $L^2({\mathbb R}^n)$ consisting of all $f\in L^2({\mathbb R}^n)$ whose Fourier 
transform $\hat f(\xi)=\int e^{-2\pi i x \cdot \xi} f(x) dx$ vanishes on  ${\mathbb R}^n\setminus K$. 
We now follow H.\,J.\,Landau.
\begin{e-definition}
A set $\Lambda\subset {\mathbb R}^n$ is a set of stable sampling for $E_K$ if there exists
a constant $C$ such that 
 
$$f\in E_K \Rightarrow \|f\|_2^2\leq C \sum_{\lambda\in \Lambda}|f(\lambda)|^2 \leqno{(5)}$$

\end{e-definition}
Then we have \cite{MM}, \cite{Mey12}
\begin{theorem}
Let $K\subset {\mathbb R}^n.$ Then simple model sets $\Lambda$ are  sets of stable sampling for $E_K$ 
whenever the density of $\Lambda$ is larger than the measure of $K.$
\end{theorem}

This theorem does not cover the limiting case where the density of $\Lambda$ equals the measure of $K.$  
In one dimension and in the periodic case an outstanding theorem by G.\,Kozma and N.\,Lev gives an answer.  
The integral part $[x]$ of  a real number  $x$ is the largest integer $k\in {\mathbb Z}$ such that $k\leq x.$  
Let $\theta>1 $ be a real number and let us define $\Lambda_{\theta}\subset {\mathbb Z}$ by 
$\Lambda_{\theta}= \{[k\theta],\,k\in {\mathbb Z}\}.$  
With these notations we have \cite{KL} 

\begin{theorem} Let us assume that $\theta>1$ is irrational. Let $S\subset {\mathbb T}$ be a finite union 
of intervals $J_m,\,1\leq m \leq M.$  We assume that the sum of the lengths of these intervals equals the 
density $\theta^{-1}$ of $\Lambda_{\theta}$ and that the length of each $J_m$ belongs to 
$\theta^{-1}{\mathbb Z}+{\mathbb Z}.$ 

\smallskip
 
Then any square summable function $f$ defined on $S$ can be uniquely written as a generalized Fourier series 
$$  
f(x)= \sum_{\lambda \in \Lambda_{\theta}} c_{\lambda}\, exp(2\pi i\lambda x) \leqno{(6)}
$$
where the frequencies $\lambda$ belong to $\Lambda_{\theta}$  and the coefficients $c_{\lambda}$ belong 
to $l^2(\Lambda_{\theta}).$ This series converges to $f$ in $L^2(S).$ 

\end{theorem}

This properties would not hold if $\Lambda$ was replaced by an ordinary lattice. Aliasing then occurs.

\section{Quasicrystals and $1$-dimensional sequences}

In quasicrystalline materials the atoms are disposed in a way which is neither periodic, as
in crystals, nor randomly disordered, as in glasses. Rather atoms follow intermediate patterns.
Furthermore the first quasicrystals described in \cite{SBGC} have a structure similar to the
($2$-dimensional) Penrose tiling, which is essentially determined by a one-dimensional binary 
sequence beginning in $0 \ 1 \ 0 \ 0 \ 1 \ 0 \ 1 \ 0 \ ...$ and known as the (binary) Fibonacci 
sequence. It was thus tempting to try and introduce ``one-dimensional'' quasicrystals as sequences
taking their values in a finite set and ``similar'' to the binary Fibonacci sequence.

\bigskip

\centerline{\epsfig{figure=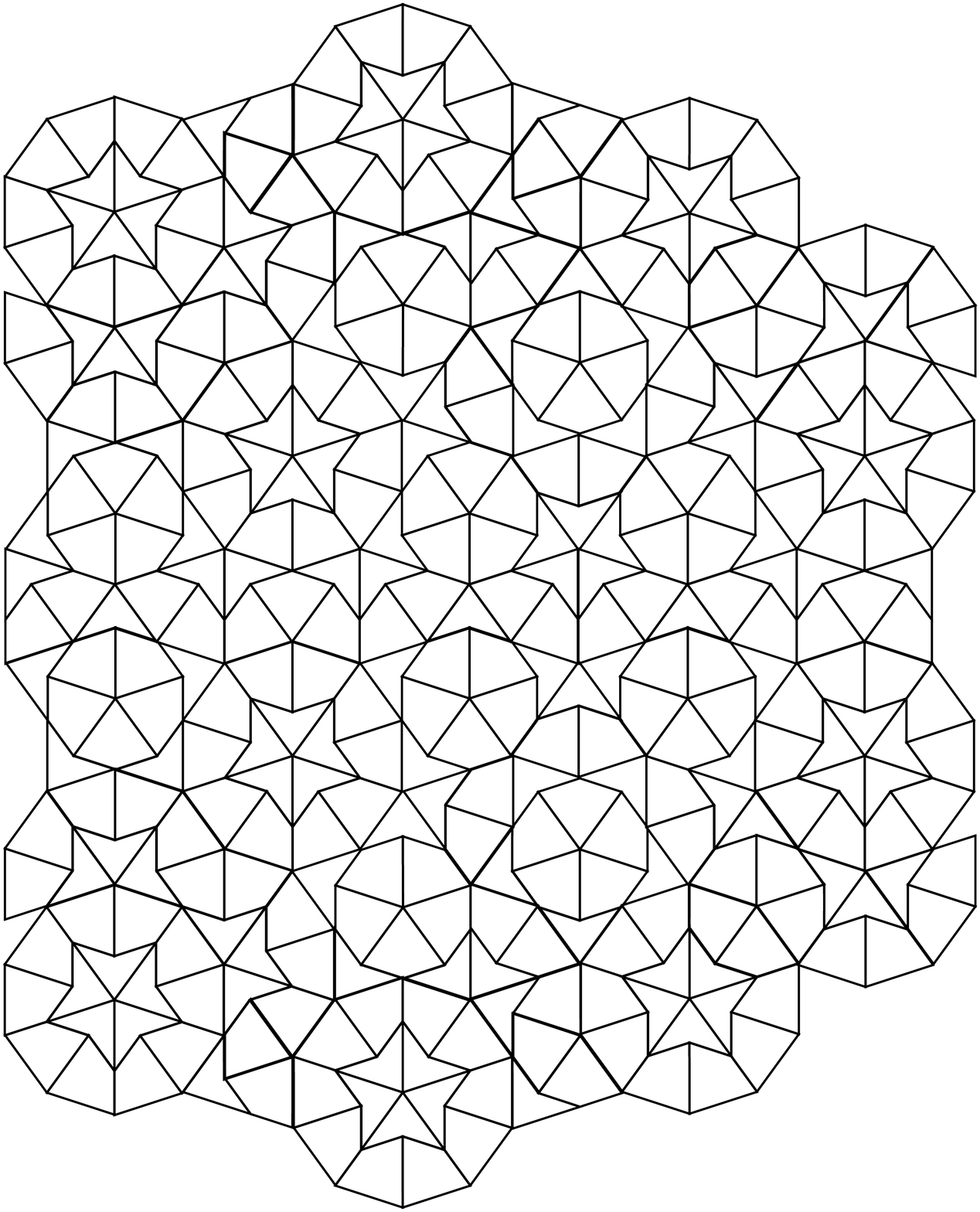,width=6cm} 
\ \ \ \ \  \epsfig{figure=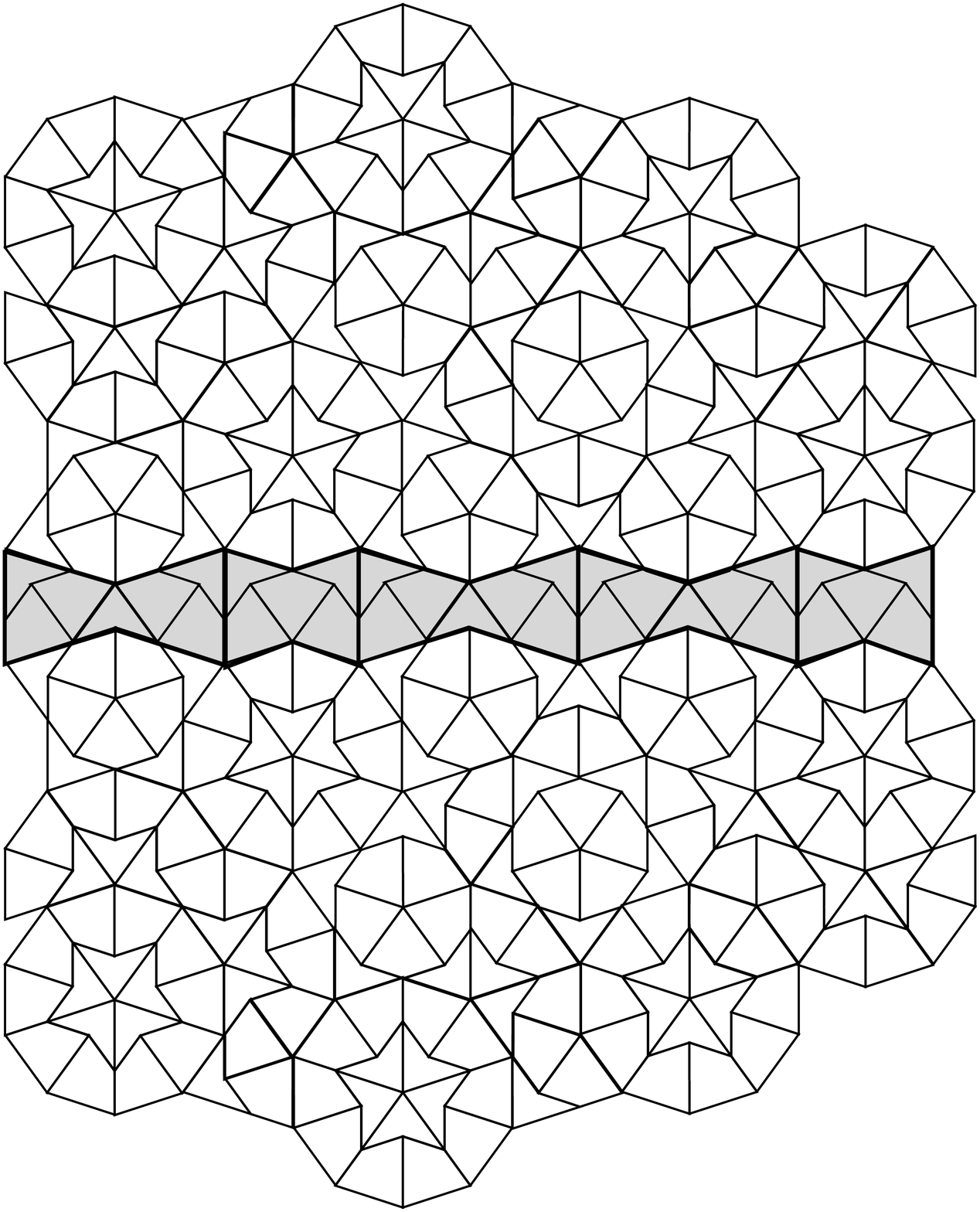,width=6cm}}

\medskip

\centerline{\scriptsize Representing on this fragment of Penrose tiling the two types of bow ties 
shaded on the right} 
\centerline{\scriptsize (long and short) by $0$ and $1$, one recognizes the beginning of the binary 
Fibonacci sequence.}

\medskip

The first papers with this approach seem to be the paper by Kohmoto, Kadanoff, and Tang \cite{KKT}
and the paper of \"Ostlund, Pandit, Rand, Schellnhuber, Siggia \cite{OPRSS}. In these two papers
the authors study a tight-binding Hamiltonian occurring in the discrete Schr\"odinger equation
where the potentials are given by a ``quasiperiodic'' sequence, i.e.,  a sequence with ``some 
order in it'' (think of a linear chain of masses and springs, where all springs are identical 
and where the masses are distributed according to some ``ordered'' sequence). The sequence considered 
in these two papers was the binary Fibonacci sequence. It happens that this sequence belongs to two 
distinct families of sequences, making both families interesting in this context: the Sturmian sequences 
on one hand, the morphic sequences on the other hand. The reader wanting to know more about discrete
Schr\"odinger operators with ``quasiperiodic potentials'', in particular in the case of Sturmian or morphic 
--also called {\it substitutional}-- potentials, can begin with the survey paper of S\"ut\H{o} and its rich 
list of references \cite{Suto}.

\subsection{Sturmian sequences}

Sturmian sequences were introduced by Morse and Hedlund in 1940 \cite{MH2}. They can be defined
either as cutting sequences or by an explicit formula (see Definition~\ref{def-sturmian} below).
Cutting the square grid ${\mathbb Z}^2$ by a straight line with irrational slope gives a sequence 
of points: define a binary sequence by choosing $0$ or $1$ according to the $n$th intersection 
being on a horizontal line or on a vertical line (we omit technicalities for when the intersection 
is both on a horizontal line and on a vertical line, which can happen only once since the slope of 
the straight line is irrational).
Of course cutting the square grid by a straight line is the same as playing billiard on a square
and coding the itinerary of the ball by $0$'s or $1$'s if it bounces on a horizontal or on a vertical
side of the square (think of folding the grid on its elementary square). These definitions can easily 
be proven equivalent to the explicit definition below. The reader interested in Sturmian sequences
can consult the second chapter of Lothaire's book \cite{Lothaire}.

\begin{e-definition}\label{def-sturmian} 
A sequence $(u_n)_{n \geq 0}$ is called {\em Sturmian} if there exist an irrational number $\alpha$
in $(0,1)$ and a real number $\rho$ such that
\begin{itemize}
\item[$*$] 
either for all $n \geq 0$, $u_n = \lfloor \alpha(n+1) + \rho \rfloor - \lfloor \alpha n + \rho \rfloor$,
\item[$*$] 
or for all $n \geq 0$, $u_n = \lceil \alpha(n+1) + \rho \rceil - \lceil \alpha n + \rho \rceil$.
\end{itemize}
where $\lfloor x \rfloor$ and $\lceil x \rceil$ stand for the lower integer part (or floor) and the upper 
integer part (or ceiling) of the real number $x$.

\medskip

Taking $\alpha = \rho = (\sqrt{5}-1)/2$ the golden ratio, one obtains the sequence
$0 \ 1 \ 0 \ 0 \ 1 \ 0 \ 1 \ 0 \ ...$ known as the (binary) Fibonacci sequence.

\end{e-definition}

\begin{remark}
Note that, by their definition, Sturmian sequences are binary sequences taking only the values
$0$ and $1$.
\end{remark}

What is remarkable is that Sturmian sequences are in some sense the ``simplest'' nonperiodic 
sequences. Namely let us define after Morse and Hedlund \cite{MH1} the (block-)complexity 
of a sequence.

\begin{e-definition}
Let ${\mathbf u} = (u_n)_{n \geq 0}$ be a sequence taking finitely many values. The 
{\em (block-)complexity} of sequence ${\mathbf u}$, denoted by $p_{\mathbf u}$, is the
function defined on ${\mathbb N} \setminus \{0\}$ by: $p_{\mathbf u}(k)$ is the number of
distinct blocks of length $k$ occurring in ${\mathbf u}$.
\end{e-definition}

\begin{remark}
If ${\mathbf u}$ takes $r$ values, then for all $k \geq 1$, one has $1 \leq p_{\mathbf u}(k) \leq r^k$.
In particular if ${\mathbf u}$ is ``random'' one should expect that all possible blocks occur in ${\mathbf u}$,
thus the complexity of ${\mathbf u}$ is maximal and equal to $r^k$. On the other hand, it is not very
difficult to prove (see \cite{MH1}) that a sequence ${\mathbf u}$ such that there exists some $k \geq 1$
with $p_{\mathbf u}(k) \leq k$ {\bf must} be periodic from some index on. In other words any nonperiodic
sequence ${\mathbf u}$ must have $p_{\mathbf u}(k) \geq k+1$ for all $k \geq 1$.

Thus the nonperiodic sequences with minimal complexity would be the sequences for which 
$p_{\mathbf u}(k) = k+1$ for all $k \geq 1$, if such sequences exist. Note that these sequences
must be binary sequences because $p_{\mathbf u}(1) = 2$. The result obtained by Morse and Hedlund 
\cite{MH2} and by Coven and Hedlund \cite{CH} is that the sequences satisfying 
$p_{\mathbf u}(k) = k+1$ for all $k \geq 1$ are {\em exactly the Sturmian sequences}.

\end{remark}

\subsection{Morphic sequences; automatic sequences}

Another property of the Fibonacci sequence is that it can be generated as follows. Start from
$0$ and replace repeatedly each $0$ by $01$, and each $1$ by $0$. This gives
$$
\begin{array}{ll}
& 0 \\
& 01 \\
& 010 \\
& 01001 \\
& 01001010 \\
& 0100101001001 \\
&... \\
\end{array}
$$
this sequence of ``words'' on the ``alphabet'' $\{0, 1\}$ converges to an infinite sequence of
$0$'s and $1$'s which is exactly the binary Fibonacci sequence. We give some formal definitions.

\begin{e-definition}
Given a finite set $A$ (also called {\em alphabet}), the free monoid generated by $A$, denoted
by $A^*$ is the set of all finite sequences (also called {\em words}) --including the empty sequence--
with values in $A$. The operation that makes $A^*$ a monoid is the {\em concatenation} of words:
it is clearly associative, and the empty word is the unit.

Given two finite sets $A$ and $B$, a {\em morphism} $\sigma$ from $A^*$ to $B^*$ is a homomorphism of 
monoids from $A^*$ to $B^*$. Is is clearly defined by its values on $A$. If the words $\sigma(a)$ all
have the same length $d$ (i.e., the same number of letters), the morphism $\sigma$ is said to be
{\em uniform} or more precisely {\em $d$-uniform}.
\end{e-definition}

\begin{e-definition}\label{fix}
Let $A$ be a finite set. Let $\sigma$ be a morphism from $A^*$ to $A^*$. Suppose that
there exists some $a \in A$ and some word $w \in A^*$ such that $\sigma(a)=aw$ and
$\sigma(w) \neq \emptyset$. Then the sequence of words $\sigma^j(a)$ converges to an 
infinite word denoted by $\sigma^{\infty}(a)$ which is said to be 
{\em an iterative fixed point} of $\sigma$.
\end{e-definition}

\begin{remark}
The topology in Definition~\ref{fix} is the topology of simple convergence, i.e., the product
topology on $A^{\mathbb N}$, where each copy of $A$ is equipped with the discrete topology.
Also $\sigma$ can be extended by continuity to $A^{\mathbb N}$, which shows that 
$\sigma(\sigma^{\infty}(a)) = \sigma^{\infty}(a)$, i.e., that $\sigma^{\infty}(a)$ is indeed
a fixed point of (the extension of) $\sigma$.
\end{remark}

\begin{example}\label{classical}
We give two classical examples.
\begin{itemize}

\item[$*$]
The Fibonacci sequence: take $A = \{0, 1\}$ and define $\sigma$ by $\sigma(0) = 01$,
$\sigma(1) = 0$. Then it can be proven that the sequence 
$\sigma^{\infty}(0) = 0 \ 1 \ 0 \ 0 \ 1 \ 0 \ 1 \ 0 \ ...$ is equal to the
Fibonacci sequence (Definition~\ref{def-sturmian}).

\item[$*$]
The Thue-Morse sequence: take $A = \{0, 1\}$ and define $\sigma$ by $\sigma(0) = 01$,
$\sigma(1) = 10$. Then $\sigma^{\infty}(0) = 0 \ 1 \ 1 \ 0 \ 1 \ 0 \ 0 \ 1 \ 1 \ ...$ which is
called the Thue-Morse (or Prouhet-Thue-Morse) sequence (see, e.g, \cite{ubiq} for a survey
of numerous properties of this sequence).

\end{itemize}

\end{example}

\begin{e-definition}\label{morphic}
Let $A$ and $B$ be two finite sets. Let $\sigma$ be a morphism from $A^*$ to $A^*$. Suppose 
that there exist some $a \in A$ and some word $w \in A^*$ such that $\sigma(a)=aw$ and
$\sigma(w) \neq \emptyset$. Let $\varphi$ be a map from $A$ to $B$. This map $\varphi$ 
can be extended (pointwise) to a map from $A^*$ to $B^*$. The sequence 
$\varphi(\sigma^{\infty}(a))$ is said to be {\em morphic}. If furthermore the morphism $\sigma$ 
is $d$-uniform, the sequence $\varphi(\sigma^{\infty}(a))$ is said to be {\em automatic} or
more precisely {\em $d$-automatic}.
\end{e-definition}

\begin{example}
The Fibonacci sequence and the Thue-Morse sequence are both morphic (see Example~\ref{classical},
and take $B=A$ and $\varphi = id$). Furthermore the Thue-Morse sequence is $2$-automatic.
\end{example}

Note that automatic sequences are, in some sense, easier to deal with than general morphic sequences. 
The reason is that the $n$th term of a $d$-automatic sequence involves the expansion of the integer 
$n$ in base $d$. This is also reflected in the number-theoretic properties of automatic sequences.
A seminal result in this direction is a theorem due to Christol \cite{Christol} and Christol, Kamae,
Mend\`es France, and Rauzy \cite{CKMFR} which states that a sequence $(a_n)_{n \geq 0}$ with values
in a finite field ${\mathbb F}_q$ is $q$-automatic if and only if the formal power series
$\sum a_n X^n$ is algebraic over the field ${\mathbb F}_q(X)$. For more about automatic sequences
see \cite{AS}, see also \cite{AMF}.

\subsection{Sturmian iterative fixed points of morphisms}

In view of the fact that the Fibonacci sequence, playing the r\^ole of a toy-model for quasicrystals, 
is both Sturmian and iterative fixed point of some morphism, one can ask for all sequences having 
this property. Interestingly enough, the question was answered without reference to or motivation from 
the theory of quasicrystals. The main result is the following theorem.

\begin{theorem}[\cite{Yasutomi}]
Let $\alpha \in (0, 1)$ and $\rho \in [0, 1)$. Define ${\mathbf u} = (u_n)_{n \geq 0}$, with
$u_n = \lfloor (n+1) \alpha + \rho \rfloor - \lfloor (n+1) \alpha + \rho \rfloor$. Then, the 
Sturmian sequence ${\mathbf u}$ is an iterative fixed point of a morphism if and only if
\begin{itemize}
\item[$*$]
$\alpha$ is quadratic, $\rho$ belongs to ${\mathbb Q}(\alpha)$;
\item[$*$]
$\alpha' > 1$, $1-\alpha' \leq \rho' \leq \alpha'$ or
$\alpha' <0$, $\alpha' \leq \rho' \leq 1-\alpha'$
\end{itemize}
where $\alpha'$ (resp.\,$\rho'$) stands for the conjugate of $\alpha$ (resp.\,$\rho$), when 
these numbers are quadratic.
\end{theorem}

\begin{remark}
A large literature addresses the question of characterizing the iterated fixed points of morphisms
that are Sturmian. Yasutomi gave the first complete answer in \cite{Yasutomi}. Other proofs were
given afterwards: let us cite in particular the proof given in \cite{BEIR} that uses {\em Rauzy fractals}
(these fractals with a flavor of number theory are a generalization of a fractal studied by G.\,Rauzy
in \cite{Rauzy}).
\end{remark}

\section{Model sets and morphic sequences}

A natural question after having seen quasicrystals as model sets, and having in mind the
morphic and automatic sequences as possible toy-models for quasicrystals, is to study
model sets that are morphic or automatic. A first idea is to demand that sequences modelling
quasicrystals be {\it repetitive} or even {\it linearly repetitive}. Note that these terms
correspond in the literature of symbolic dynamics to what is called {\it uniformly recurrent} 
resp.\,{\it linearly uniformly recurrent} sequences. Recall that a sequence is said to be repetitive
if for each $k >0$, there exists an integer $M(k)$ such that any block of length $M(k)$ of the
sequence contains at least one copy of every block of length $k$ occurring in the sequence.
The sequence is called linearly repetitive if furthermore one can choose $M(k) = O(k)$.
These notions are extended to two dimensions in the nice paper by Lagarias and Pleasants
\cite{LP}.

\bigskip

Another idea is, since automatic sequences are easier to generalize to several dimensions than 
morphic sequences, to try to characterize $2$-dimensional sequences that are automatic. Not many 
papers were devoted to this study. We cite one article by Barb\'e and von Haeseler \cite{BH} 
giving a necessary and sufficient condition for a $2$-dimensional automatic sequence to be Delone. 
The condition is too technical to be given here. We just note that a $2$-dimensional generalization 
of the Thue-Morse sequence is Delone, while the authors give other examples of automatic sequences 
that are or are not Delone.

\end{document}